\documentclass[a4paper,12pt]{article}

\pdfoutput=1
\usepackage{amsmath}
\usepackage{amssymb}
\usepackage{amsfonts}
\usepackage{mathrsfs}
\usepackage{bbm}
\usepackage{graphicx,subfigure,booktabs}
\usepackage[numbers,sort&compress]{natbib}
\usepackage{verbatim}
\usepackage{color}
\usepackage{ulem}
\usepackage{setspace}
\usepackage{multirow}
\usepackage{url}
\usepackage{adjustbox}
\usepackage[utf8]{inputenc}
\usepackage{makecell}
\usepackage[colorlinks,
linkcolor=black,
filecolor=black,
anchorcolor=black,
urlcolor=black,
citecolor=blue,
bookmarks=false,
]{hyperref}

\usepackage[hyphenbreaks]{breakurl}

\usepackage{fancyhdr}

\numberwithin{equation}{section}

\newlength{\dinwidth}
\newlength{\dinmargin}
\allowdisplaybreaks

\begin{document}



\title{Nonperturbative effects in neutrino magnetic moments}

\author{Feng-Zhi Chen${}^{1,2,}$\footnote{chenfzh25@mail.sysu.edu.cn}\,\,,
Min-Di Zheng${}^{1,}$\footnote{zhengmd5@mail.sysu.edu.cn}\,\,,
and Hong-Hao Zhang${}^{1,}$\footnote{zhh98@mail.sysu.edu.cn}\\[12pt]
\small ${}^{1}$ School of Physics, Sun Yat-Sen University, Guangzhou 510275, China\\[-0.2cm]
\small ${}^{2}$ Key Laboratory of Quark and Lepton Physics~(MOE),\\[-0.2cm]
\small Central China Normal University, Wuhan 430079,China}
\date{}
\maketitle
\vspace{-0.2cm}

\begin{abstract}
{\noindent}In this paper, we calculate the QCD nonperturbative contributions of the neutrino-quark tensor operators to the neutrino magnetic moments by matching onto the chiral perturbation theory at low energies. These nonperturbative contributions can be compared to the perturbative ones, which are induced from one-loop mixing when performing the renormalization group evolutions from $\mu=m_W$ down to $\mu=2~\mathrm{GeV}$. We then constrain the dipole and tensor Wilson coefficients of the low-energy neutrino effective field theory (LNEFT) separately from the neutrino-electron scattering with Borexino data and coherent elastic neutrino-nucleus scattering (CE$\nu$NS) with COHERENT data to show the competition between these two contributions, at the renormalization scales $\mu=2~\mathrm{GeV}$ and $\mu=m_W$ in the $\overline{\mathrm{MS}}$ scheme. In the neutrino-electron scattering, it is found that the nonperturbative contributions dominate for the coefficients involving
up and down quarks, while they are expected to be of the same order of magnitude as the perturbative contributions for the coefficients involving strange quark. As for constraints
in the CE$\nu$NS, the tensor operators can contribute to the process through either direct or indirect way. As a result, the indirect contributions including nonperturbative and perturbative parts for all couplings become negligible in comparison with the direct ones. As the nonperturbative contributions crucially depend on the value of $c_T$, its inputs will affect the extraction of limits on the tensor  LNEFT Wilson coefficients. We compute the upper bounds on these coefficients with $c_T$ quoting from the model and lattice estimates.
\end{abstract}

\newpage
\section{Introduction}
\label{sec:introduction}

The neutrino magnetic moment (NMM) plays an important role in the exploration of new physics (NP) beyond the Standard Model (SM). In the minimal extension of the SM with three right-handed neutrinos, the NMM can occur at loop level with external photon attaching to the charged leptons in the loops, and its magnitude is depending on the input of the neutrino mass $m_\nu$. Given $m_\nu\leq1~\mathrm{eV}$, the NMM is predicted to be less than $\mathrm{a~few}\times10^{-19}\mu_B$~\cite{Fujikawa:1980yx,Pal:1981rm,Shrock:1982sc,Dvornikov:2003js,Dvornikov:2004sj}, with $\mu_B=e/2m_e$ standing for the Bohr magneton. This is far below the best upper bounds from the terrestrial experiments GEMMA (based on reactor neutrinos sources)~\cite{Beda:2012zz} and Borexino (based on solar neutrinos sources)~\cite{Borexino:2017fbd}, which are of order $\mathcal{O}(10^{-11})\mu_B$. Nevertheless, the NMMs are not necessary to be proportional to the neutrino mass in the presence of NP, so their magnitudes may be much larger that they can reach the detection sensitivity of current or future experiments. For a comprehensive review on this regard, one is referred to Ref.~\cite{Giunti:2014ixa}.

As the processes relevant to the NMMs usually occur at the energies that are far below the electroweak scale $\Lambda_\mathrm{EW}$, a general model-independent  treatment is to employ the low energy effective field theory (LEFT)~\cite{Jenkins:2017jig,Jenkins:2017dyc}, which respects to $SU(3)_C\times U(1)_{em}$ gauge symmetries, to describe the NP impacts from higher scales. However, since there only contains left-handed neutrinos $\nu_L$ in the LEFT, one can merely construct lepton number violating (LNV) operators, which are chirality flipping, for left-handed Majorana neutrinos transition magnetic moments. To include also lepton number conserving (LNC) operators for Dirac NMMs, one has to extend the LEFT with right-handed neutrinos $N_R$, the resulting effective field theory of which is now well known as LNEFT (see e.g. Refs.~\cite{Chala:2020vqp,Li:2020lba}). If we further assume that the NP is from the scale well above $\Lambda_\mathrm{EW}$, then the SM effective field theory (SMEFT)~\cite{Buchmuller:1985jz,Grzadkowski:2010es,Lehman:2014jma,Brivio:2017vri} augmented with right-handed neutrinos (SMNEFT)~\cite{delAguila:2008ir,Aparici:2009fh,Bhattacharya:2015vja,Liao:2016hru,Liao:2016qyd,Bischer:2019ttk}, which respects to  the SM gauge symmetries $SU(3)_C\times SU(2)_L\times U(1)_{Y}$, will provide an adequate description to  these interactions at the domain from $\Lambda_\mathrm{EW}$ up to NP scale $\Lambda_\mathrm{NP}$. After matching the LNEFT operators onto the SMNEFT operators at $\Lambda_\mathrm{EW}$ and taking into account the renormalization group (RG) running effects, one can then translate the bounds obtained from low-energy processes into the constraints at $\Lambda_\mathrm{NP}$. 

\begin{figure}[t!]
	\centering
	\includegraphics[width=0.27\textwidth]{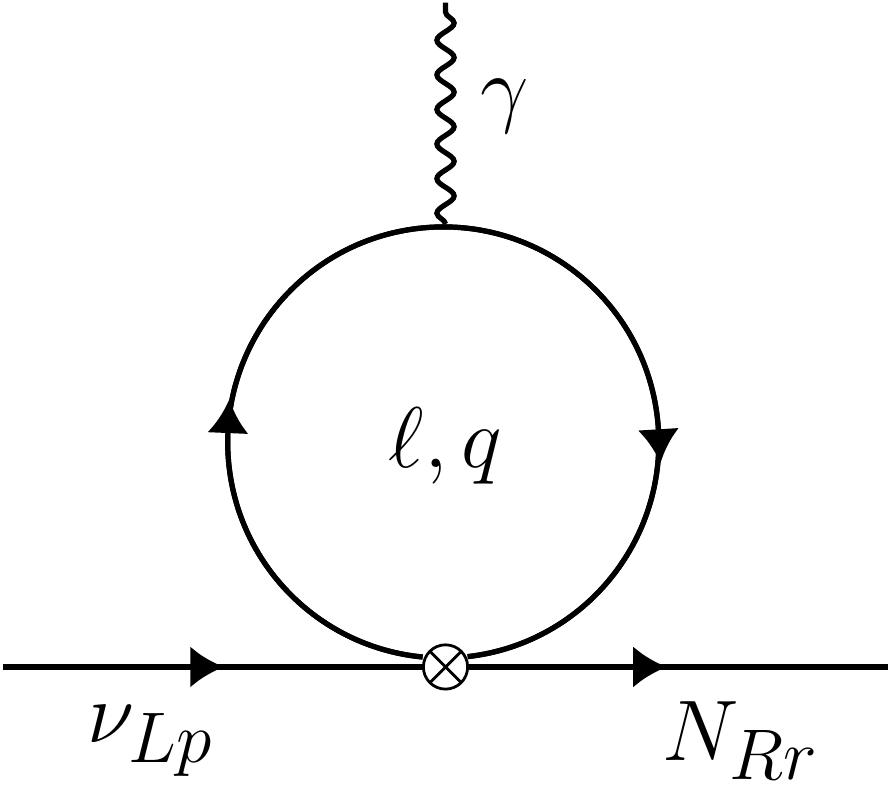}
	\caption{\small Feynman diagram describing the neutrino-lepton or neutrino-quark tensor interacting operators (crossed circle) contribute to the Diac NMMs via one-loop mixing. Similar diagram describing the Majorana transition NMMs can be obtained by replacing $\nu_L$($N_R$) with $N_R^c$($\nu_L^c$), meanwhile the off-diagonal flavor condition $p\neq r$ should be understood due to the charge-conjugation, parity-transformation, and time-reversal (CPT) conservation.\label{fig:NMM}}
\end{figure}

In the LNEFT, the leading operators that contribute to the NMMs are the dimension-5 dipole operators. In addition to the dipole operators, there are a subset of dimension-6 tensor operators, which describe the neutrino-lepton and neutrino-quark interactions, can also contribute to the NMMs. When working at the renormalization scale $\mu=2~\mathrm{GeV}$ at which perturbation theory is still valid, the heavy quarks can be integrated out, leaving the leptons together with three light quarks $q=u,d,s$ as the active degrees of freedom. In the perturbative scales ($\mu\geq2~\mathrm{GeV}$), the tensor operators can contribute to the NMMs via one-loop Feynman diagram, see figure~\ref{fig:NMM}. Such a diagram can stem from, e.g., the minimal left-right symmetric model~\cite{Pati:1974yy,Mohapatra:1974gc,Senjanovic:1975rk} or the $\tilde{R}_2$ scalar leptoquark model~\cite{Dorsner:2016wpm}, see App.~\ref{app:UV} for more details. As for energies below $2~\mathrm{GeV}$, the nonperturbative effects induced from the neutrino-quark interactions will be of importance, and are needed to be taken into account carefully. 

In this work, we shall follow a similar procedure which computes the nonperturbative effects in the charged lepton flavor violating (CLFV) process $\mu\to e\gamma$ in Ref.~\cite{Dekens:2018pbu}, the nonperturbative effects in the NMMs can be obtained by matching the dimension-6 LNEFT tensor operators onto the operators of chiral perturbation theory ($\chi$PT) with tensor external sources~\cite{Weinberg:1968de,Gasser:1983yg,Gasser:1984gg,Cata:2007ns}. We find that the nonperturbative contributions from the tensor operators to the dipole operators are of order $L_{i\gamma}\sim e(F_\pi^2/\Lambda_\chi)L_{iq}^{T,AA}$, where $i\in(\nu N,\nu,N)$ and $A\in(L,R)$. These results can be compared to the perturbative contributions, which are of order $L_{i\gamma}\sim e m_q/(16\pi^2)L_{iq}^{T,AA}$, and are  obtained from one-loop mixing when taking into account the RG running from $\mu=\Lambda_\mathrm{EW}$ down to $\mu=2~\mathrm{GeV}$. To show the competition between these two effects, we will constrain the relevant LNEFT Wilson coefficients separately from the neutrino-electron scattering with the Borexino data~\cite{Borexino:2017fbd} and CE$\nu$NS with the COHERENT data~\cite{COHERENT:2017ipa}. For more details, readers are referred to Sec.~\ref{sec:match} and~\ref{sec:numerical}.

This paper is outlined as follows. In Sec.~\ref{sec:match}, we will introduce the relevant  operators which have contributions to the NMMs in LNEFT as well as $\chi$PT, and then match the corresponding Wilson coefficients of the two theories at $\mu=2~\mathrm{GeV}$. The numerical bounds on the Wilson coefficients of LNEFT dipole and tensor operators at renormalization scales $\mu=2~\mathrm{GeV}$ and $\mu=m_W$ will be given in Sec.~\ref{sec:numerical}. Our conclusions are drawn in Sec.~\ref{sec:conclusions}.

\section{Matching LNEFT to $\chi$PT}\label{sec:match}

In this section we will briefly introduce the building blocks that are required for this work. We will first list the relevant LNEFT operators; the RG equations for the respective Wilson coefficients are also given. The LNEFT operators should be matched onto the SMNEFT operators if NP is from the scale well above the electroweak. Then the nonperturbative effects can be computed by matching the LNEFT operators onto the low-energy operators of $\chi$PT.

\subsection{LNEFT and RG evolutions}

The LNEFT Lagrangian, which respects to $SU(3)_C\times U(1)_{em}$ gauge symmetries, can be written schematically as~\cite{Jenkins:2017jig,Jenkins:2017dyc,Chala:2020vqp,Li:2020lba}
\begin{align}\label{eq:LNEFT}
\mathcal{L}_\mathrm{LNEFT}=\mathcal{L}_{d\leq4}+\sum_{d\geq5}\sum_iL_i^{(d)}\mathcal{O}_i^{(d)}\,,
\end{align}
where the first term in the right side consists of the normal QED and QCD terms of the SM light particles (the heavy particles $h$, $W$, $Z$, and $t$ have been integrated out) as well as the Majorana kinetic and mass terms of left- and right-handed neutrinos, whereas the second term includes the dimension-$d$ ($d\geq5$) operator $\mathcal{O}_i^{(d)}$ with the corresponding Wilson coefficient $L_i^{(d)}$. When encoding the right-handed neutrinos in LNEFT, we have assumed that they are light particles with masses well below $\Lambda_\mathrm{EW}$. In principle, the flavors of the right-handed neutrinos can be arbitrary, albeit many experiments implicitly suppose that there are three, with one flavor for each generation. As we are interested in the nonperturbative effects in the NMMs, this would involve the dimension-5 dipole operators as well as the dimension-6 neutrino-quark interacting operators. The relevant effective operators can be classified into two subsets, i.e., the LNC and LNV operators~\cite{Chala:2020vqp,Li:2020lba}. Adopting the notations used in Ref.~\cite{Jenkins:2017jig}, we collect the relevant LNC ($|\Delta L|=0$) LNEFT operators in table~\ref{tab:LNC}, and list the LNV ($|\Delta L|=2$) operators in table~\ref{tab:LNV}. For the notations of the Wilson coefficients, we use the same subscripts and superscripts as the operators, for instance $L_{\substack{\nu u\\prwt}}^{V,LL}$ together with $\mathcal{O}_{\substack{\nu u\\prwt}}^{V,LL}$, where $p,r,w,t$ are quark or neutrino flavor indices. The charge conjugation of a left-handed neutrino $\nu_L^c=\mathcal{C}\bar{\nu}_L^T$ ($\mathcal{C}=i\gamma_2\gamma_0$) is a right-handed field, while the charge conjugation of a right-handed neutrino $N_R^c=\mathcal{C}\bar{N}_R^T$ is a left-handed field. The operators grouped in each tables are classified according to the chirality $L$ and $R$ of the fermion bilinears. The complete set of the operators in other notations can also be found in Refs.~\cite{Chala:2020vqp,Li:2020lba}.

\begin{table}[h!]
\small
\centering
\resizebox{\linewidth}{!}{
\renewcommand{\arraystretch}{1.5}
\begin{tabular}[t]{c|c}
    \multicolumn{2}{c}{\boldmath$(\overline{L}R)X+\mathrm{h.c}$} \\
\hline
$\mathcal{O}_{\nu N\gamma}$ & $(\bar \nu_{Lp} \sigma^{\mu\nu} N_{Rr})F_{\mu\nu}$\\
	\multicolumn{2}{c}{\boldmath$(\overline{L} L)(\overline{L} L)$} \\
	\hline
	$\mathcal{O}_{\nu u}^{V,LL}$ & $(\bar \nu_{Lp} \gamma^\mu \nu_{Lr})(\bar u_{Lw} \gamma_\mu u_{Lt})$ \\
	$\mathcal{O}_{\nu d}^{V,LL}$ & $(\bar \nu_{Lp} \gamma^\mu \nu_{Lr})(\bar d_{Lw} \gamma_\mu d_{Lt})$ \\
	\multicolumn{2}{c}{\boldmath$(\overline{R} R)(\overline{R} R)$} \\
	\hline
	$\mathcal{O}_{N u}^{V,RR}$ & $(\bar N_{Rp} \gamma^\mu N_{Rr})(\bar u_{Rw} \gamma_\mu u_{Rt})$ \\
	$\mathcal{O}_{N d}^{V,RR}$ & $(\bar N_{Rp} \gamma^\mu N_{Rr})(\bar d_{Rw} \gamma_\mu d_{Rt})$ \\
\end{tabular}
\begin{tabular}[t]{c|c}
\multicolumn{2}{c}{\boldmath$(\overline{L} L)(\overline{R} R)$} \\
	\hline
	$\mathcal{O}_{\nu u}^{V,LR}$ & $(\bar \nu_{Lp} \gamma^\mu \nu_{Lr})(\bar u_{Rw} \gamma_\mu u_{Rt})$ \\
	$\mathcal{O}_{\nu d}^{V,LR}$ & $(\bar \nu_{Lp} \gamma^\mu \nu_{Lr})(\bar d_{Rw} \gamma_\mu d_{Rt})$ \\
	$\mathcal{O}_{uN}^{V,LR}$    & $(\bar u_{Lp} \gamma^\mu u_{Lr})(\bar N_{Rw} \gamma_\mu N_{Rt})$   \\
	$\mathcal{O}_{dN}^{V,LR}$ & $(\bar d_{Lp} \gamma^\mu d_{Lr})(\bar N_{Rw} \gamma_\mu N_{Rt})$ \\
\end{tabular}
\begin{tabular}[t]{c|c}
\multicolumn{2}{c}{\boldmath$(\overline{L} R)(\overline{L} R) + \mathrm{h.c.}$} \\
	\hline
	$\mathcal{O}_{\nu Nu}^{S,RR}$ & $(\bar \nu_{Lp} N_{Rr})(\bar u_{Lw} u_{Rt})$ \\
	$\mathcal{O}_{\nu Nd}^{S,RR}$ & $(\bar \nu_{Lp} N_{Rr})(\bar d_{Lw} d_{Rt})$ \\
	$\mathcal{O}_{\nu Nu}^{T,RR}$ & $(\bar \nu_{Lp} \sigma^{\mu\nu} N_{Rr})(\bar u_{Lw} \sigma_{\mu\nu} u_{Rt})$ \\
	$\mathcal{O}_{\nu Nd}^{T,RR}$ & $(\bar \nu_{Lp} \sigma^{\mu\nu} N_{Rr})(\bar d_{Lw} \sigma_{\mu\nu} d_{Rt})$ \\
	\multicolumn{2}{c}{\boldmath$(\overline{L} R)(\overline{R} L) + \mathrm{h.c.}$} \\
	\hline
	$\mathcal{O}_{\nu Nu}^{S,RL}$ & $(\bar \nu_{Lp} N_{Rr})(\bar u_{Rw} u_{Lt})$ \\
	$\mathcal{O}_{\nu Nd}^{S,RL}$ & $(\bar \nu_{Lp} N_{Rr})(\bar d_{Rw} d_{Lt})$\\
\end{tabular}}
\caption{The LNC ($|\Delta L|=0$) LNEFT operators including dimension-5 neutrino dipole operators and dimension-6 neutrino-quark interacting operators.}\label{tab:LNC}
\end{table}
\begin{table}[h!]
\small
\centering
\resizebox{\linewidth}{!}{
\renewcommand{\arraystretch}{1.51}
\begin{tabular}[t]{c|c}
    \multicolumn{2}{c}{\boldmath$(\overline{R}L)X+ \mathrm{h.c.}$} \\
	\hline
	$\mathcal{O}_{\nu \gamma}$ & $(\bar \nu^c_{Lp} \sigma^{\mu\nu} \nu_{Lr})F_{\mu\nu}$ \\
    \multicolumn{2}{c}{\boldmath$(\overline{L}R)X+ \mathrm{h.c.}$} \\
	\hline
	$\mathcal{O}_{N \gamma}$ & $(\bar{N}_{Rp}^c \sigma^{\mu\nu} N_{Rr})F_{\mu\nu}$ \\
	\multicolumn{2}{c}{\boldmath$(\overline{R} R)(\overline{L} L)+ \mathrm{h.c.}$} \\
	\hline
	$\mathcal{O}_{\nu Nu}^{V,RL}$ & $(\bar \nu^c_{Lp} \gamma_\mu N_{Rr})(\bar u_{Lw} \gamma^\mu u_{Lt})$ \\
	$\mathcal{O}_{\nu Nd}^{V,RL}$ & $(\bar \nu^c_{Lp} \gamma_\mu N_{Rr})(\bar d_{Lw} \gamma^\mu d_{Lt})$ \\
\multicolumn{2}{c}{\boldmath$(\overline{R} R)(\overline{R} R)+ \mathrm{h.c.}$} \\
	\hline
	$\mathcal{O}_{\nu Nu}^{V,RR}$ & $(\bar \nu^c_{Lp} \gamma_\mu N_{Rr})(\bar u_{Rw} \gamma^\mu u_{Rt})$ \\
	$\mathcal{O}_{\nu Nd}^{V,RR}$ & $(\bar \nu^c_{Lp} \gamma_\mu N_{Rr})(\bar d_{Rw} \gamma^\mu d_{Rt})$ \\
\end{tabular}
\begin{tabular}[t]{c|c}
\multicolumn{2}{c}{\boldmath$(\overline{R} L)(\overline{R} L) + \mathrm{h.c.}$} \\
	\hline
	$\mathcal{O}_{\nu u}^{S,LL}$ & $(\bar \nu^c_{Lp} \nu_{Lr})(\bar u_{Rw} u_{Lt})$ \\
	$\mathcal{O}_{\nu d}^{S,LL}$ & $(\bar \nu^c_{Lp} \nu_{Lr})(\bar d_{Rw} d_{Lt})$ \\
	$\mathcal{O}_{\nu u}^{T,LL}$ & $(\bar \nu^c_{Lp} \sigma^{\mu\nu} \nu_{Lr})(\bar u_{Rw} \sigma_{\mu\nu} u_{Lt})$ \\
	$\mathcal{O}_{\nu d}^{T,LL}$ & $(\bar \nu^c_{Lp} \sigma^{\mu\nu} \nu_{Lr})(\bar d_{Rw} \sigma_{\mu\nu} d_{Lt})$ \\
\multicolumn{2}{c}{\boldmath$(\overline{R} L)(\overline{L} R) + \mathrm{h.c.}$} \\
	\hline
	$\mathcal{O}_{\nu u}^{S,LR}$ & $(\bar \nu^c_{Lp} \nu_{Lr})(\bar u_{Lw} u_{Rt})$ \\
	$\mathcal{O}_{\nu d}^{S,LR}$ & $(\bar \nu^c_{Lp} \nu_{Lr})(\bar d_{Lw} d_{Rt})$ \\
	$\mathcal{O}_{uN }^{S,LR}$ & $(\bar u_{Rp} u_{Lr})(\bar N^c_{Rw} N_{Rt})$ \\
	$\mathcal{O}_{dN }^{S,LR}$ & $(\bar d_{Rp} d_{Lr})(\bar N^c_{Rw} N_{Rt})$ \\
\end{tabular}
\begin{tabular}[t]{c|c} 
	\multicolumn{2}{c}{\boldmath$(\overline{L} R)(\overline{L} R) + \mathrm{h.c.}$} \\
	\hline
	$\mathcal{O}_{N u}^{S,RR}$ & $(\bar N^c_{Rp} N_{Rr})(\bar u_{Lw} u_{Rt})$ \\
	$\mathcal{O}_{N d}^{S,RR}$ & $(\bar N^c_{Rp} N_{Rr})(\bar d_{Lw} d_{Rt})$ \\
	$\mathcal{O}_{N u}^{T,RR}$ & $(\bar N^c_{Rp} \sigma^{\mu\nu} N_{Rr})(\bar u_{Lw} \sigma_{\mu\nu} u_{Rt})$ \\
	$\mathcal{O}_{N d}^{T,RR}$ & $(\bar N^c_{Rp} \sigma^{\mu\nu} N_{Rr})(\bar d_{Lw} \sigma_{\mu\nu} d_{Rt})$ \\
\end{tabular}}
\caption{The LNV ($|\Delta L|=2$) LNEFT operators including dimension-5 neutrino dipole operators and dimension-6 neutrino-quark interacting operators.}\label{tab:LNV}
\end{table}

As the perturbative property of QCD is valid in the domain from $\mu=2~\mathrm{GeV}$ up to $\Lambda_\mathrm{EW}$, to compute the perturbative contributions to NMMs via figure~\ref{fig:NMM}, one has to take into account the running and mixing effects of the relevant Wilson coefficients, which are governed by the RG equations. Working in the $\overline{\mathrm{MS}}$ renormalization scheme, the RG equation for the Wilson coefficients can be written schematically as
\begin{align}\label{eq:RGE}
\dot{L}(\mu)\equiv16\pi^2\mu\frac{d L(\mu)}{d\mu}=\hat{\gamma}(\mu)L(\mu)\,, 
\end{align}
with
\begin{align}
L(\mu)=(L_1(\mu),L_2(\mu),\cdots)^T\,.
\end{align}
Here, $\mu$ is the renormalization scale, and $\hat{\gamma}(\mu)$ is the anomalous dimension matrix which is function of QCD and QED gauge couplings. When running the Wilson coefficients from a high scale $\mu_h$ down to a low scale $\mu_l$, the approximate analytic solution of RG equation in Eq.~\eqref{eq:RGE} reads~\cite{Davidson:2016edt,Cirigliano:2017azj}
\begin{align}\label{eq:solution}
L_I(\mu_l)=L_J(\mu_h)\lambda^{a_J}\left(\delta_{JI}-\frac{\alpha}{4\pi}\tilde{\gamma}_{JI}^e\log\frac{\mu_h}{\mu_l}\right)\,,
\end{align}
where $\alpha=e^2/4\pi$ is the fine structure constant, and 
\begin{align}
\lambda=\frac{\alpha_s(\mu_h)}{\alpha_s(\mu_l)}\,,\qquad a_J=\frac{\gamma^s_J}{2\beta_0}\,,\qquad \tilde{\gamma}_{JI}^e=\gamma_{JI}^ef_{JI}\,,
\end{align}
with $\beta_0=(11N_C-2n_f)/3$ ($N_C=3$ is the color number and $n_f$ is the active quark number), $\gamma_J^s$ and $\gamma_{JI}^e$ denote respectively the QCD and QED anomalous dimensions, and the definition of function $f_{JI}$ in the reduced QED anomalous dimension $\tilde{\gamma}_{JI}^e$ is
\begin{align}
f_{JI}=\frac{1}{1+a_J-a_I}\frac{\lambda^{a_I-a_J}-\lambda}{1-\lambda}\,.
\end{align}

Explicitly, in App.~\ref{app:RGE} we list the RG equations for Wilson coefficients of both LNC and LNV dimension-5 dipole operators as well as dimension-6 neutrino-quark interacting operators that we will use in Eqs.~\eqref{eq:LNCRGE} and~\eqref{eq:LNVRGE}. Note that the vector operators are not renormalized at one-loop level due to the QED and QCD Ward identities. It is clear from Eqs.~\eqref{eq:LNCRGE} and~\eqref{eq:LNVRGE} that the tensor operators can mix into the dipole operators when performing the RG evolutions. Besides, at one-loop level there is no mixing occurs between scalar and tensor operators for neutrino-quark interactions, which is contrary to the large mixing effects of the same type operators for charged lepton-quark interactions~\cite{Cirigliano:2017azj,Dekens:2018pbu}.

If we further assume that the NP is from the scale well above $\Lambda_\mathrm{EW}$, then the physics at the domain from $\Lambda_\mathrm{EW}$ up to $\Lambda_\mathrm{NP}$ can be described by the SMNEFT Lagrangian~\cite{delAguila:2008ir,Aparici:2009fh,Bhattacharya:2015vja,Liao:2016hru,Liao:2016qyd,Bischer:2019ttk}
\begin{align}\label{eq:SMNEFT}
\mathcal{L}_\mathrm{SMNEFT}=\mathcal{L}_\mathrm{SM+N}+\sum_{d\geq5}\sum_iC_i^{(d)}\mathcal{Q}_i^{(d)}\,,
\end{align}
where $\mathcal{L}_\mathrm{SM+N}$ is the SM Lagrangian extended with right-handed neutrinos $N_R$, whereas $\mathcal{Q}_i^{(d)}$ and $C_i^{(d)}$ are the dimension-$d$ ($d\geq5$) operators and their respective Wilson coefficients. The SMNEFT Lagrangian in Eq.~\eqref{eq:SMNEFT} respects to $SU(3)_C\times SU(2)_L\times U(1)_Y$ gauge symmetries. In order to translate the bounds from the low-energy physical processes into the constraints at $\Lambda_\mathrm{NP}$, we need four steps: matching at experimental scale $\Lambda_\mathrm{Exp}$, running up to $\Lambda_\mathrm{EW}$, matching at $\Lambda_\mathrm{EW}$, then running up to $\Lambda_\mathrm{NP}$. The RG equations of the SMNEFT Wilson coefficients with the one-loop anomalous dimensions from electroweak gauge as well as Yukawa corrections have been calculated in Refs.~\cite{Bell:2005kz,Chala:2020pbn,Han:2020pff,Datta:2020ocb,Datta:2021akg}. As for operator matching of LNEFT and SMNEFT at $\Lambda_\mathrm{EW}$, readers are referred to Refs.~\cite{Jenkins:2017jig,Chala:2020vqp,Han:2020pff,Li:2020lba}. 

\subsection{$\chi$PT and operators matching}

The $\chi$PT is valid below the chiral symmetries breaking scale $\Lambda_\chi\sim4\pi F_\pi$~\cite{Manohar:1983md}, with the physical pion decay constant $F_\pi=92.3(1)~\mathrm{MeV}$. For energies well below $\Lambda_\chi$, the light quark bilinears can be matched onto the $\chi$PT operators.  Such a matching has already been carried out in the charged lepton sector~\cite{Dekens:2018pbu}, as well as the neutrino sector~\cite{Li:2019fhz}, with the external fields derived from the LEFT. In this work, we will follow a similar procedure done in Ref.~\cite{Dekens:2018pbu} for the computation of nonperturbative effects in the CLFV process $\mu\to e\gamma$, and calculate the nonperturbative effects in the NMMs with the external sources induced from the LNEFT as follows.

We start with the massless QCD Lagrangian extended with quark bilinears coupling to external sources~\cite{Gasser:1983yg,Gasser:1984gg,Cata:2007ns,Dekens:2018pbu}~\footnote{We adopt the notations used in Ref.~\cite{Dekens:2018pbu}.}
\begin{align}\label{eq:QCD}
\mathcal{L}&=\mathcal{L}_\mathrm{QCD}^0+\mathcal{L}_\mathrm{ext}\nonumber\\
&=\mathcal{L}_\mathrm{QCD}^0+\bar{q}_L\gamma^\mu l_\mu q_L+\bar{q}_R\gamma^\mu r_\mu q_R+\bar{q}_LSq_R+\bar{q}_RS^\dagger q_L+\bar{q}_L\sigma^{\mu\nu} t_{\mu\nu} q_R+\bar{q}_R\sigma_{\mu\nu} t_{\mu\nu}^\dagger q_L\,,
\end{align}
where $q=(u,d,s)^T$, and $l_\mu$, $r_\mu$, $S$, and $t_{\mu\nu}$ are $3\times3$ Hermitian matrices in flavor space, denoting the left-handed, right-handed, scalar, and tensor external sources, respectively. The corresponding $\chi$PT Lagrangian, which describes the strong interactions of the dynamical meson fields $(\pi,K,\eta)$ and their couplings to the external sources, have been worked out in Refs.~\cite{Gasser:1983yg,Gasser:1984gg,Cata:2007ns}. The $\chi$PT is based on the global $SU(3)_L\times SU(3)_R$ flavor symmetries spontaneously broken down to $SU(3)_V$, so their interactions can be completely described by the Goldstone dynamics. The nonlinear realization of the theory is embedding the Goldstone octet into the matrix representation $U$, with
\begin{align}
U=\mathrm{exp}\left(i\frac{\Phi}{F_0}\right)\,,\qquad \Phi=\lambda^a\phi^a=\begin{pmatrix}
\pi^0+\frac{1}{\sqrt{3}}\eta & \sqrt{2}\pi^+ &\sqrt{2}K^+\\
\sqrt{2}\pi^- & -\pi^0+\frac{1}{\sqrt{3}}\eta & \sqrt{2}K^0\\
\sqrt{2}K^- &\sqrt{2}\bar{K}^0 & -\frac{2}{\sqrt{3}}\eta\\
\end{pmatrix}\,,
\end{align}
where $F_0$ is the pion decay constant in the chiral limit, $\phi^a$ are the Goldstone bosons, and $\lambda^a$ ($a=1,\cdots,8$) are the Gell-Mann matrices satisfying the trace relation $\mathrm{Tr}(\lambda_a\lambda_b)=2\delta_{ab}$. The $U$ matrix transforms as $U\to RUL^\dagger$ under the chiral symmetries, with $L(R)\in SU(3)_{L(R)}$. By making use of the external field method~\cite{Gasser:1983yg}, the neutrino bilinears together with the accompanied Wilson coefficients in Eq.~\eqref{eq:LNEFT}, can be treated as external sources (which behave as spurion fields~\cite{Dekens:2018pbu}) that follow the chiral symmetries and are endowed with the following chiral power counting 
\begin{align}\label{eq:counting}
l_\mu\sim \mathcal{O}(p)\,,\quad r_\mu\sim \mathcal{O}(p)\,,\quad S\sim \mathcal{O}(p^2)\,,\quad t_{\mu\nu}\sim \mathcal{O}(p^2)\,,
\end{align}
in which they can be organized orderly into the $\chi$PT Lagrangian. Then the Green function of quark bilinears can be obtained by taking the functional derivatives with respect to the external sources.

The external sources may come from either the SM or NP, so we can explicitly split them into two parts~\cite{Dekens:2018pbu}:
\begin{align}
l_\mu\mapsto l_\mu+\tilde{l}_\mu\,,\quad r_\mu\mapsto r_\mu+\tilde{r}_\mu\,,\quad S\mapsto S+\tilde{S}\,,\quad t_{\mu\nu}\mapsto t_{\mu\nu}+\tilde{t}_{\mu\nu}\,,
\end{align}
where $S$, $l_\mu$, $r_\mu$, and $t_{\mu\nu}$ in the right side of each pieces describe the quark mass matrix and the couplings to electromagnetic field:
\begin{align}\label{eq:split}
S\mapsto -M^\dagger\,,\quad l_\mu\mapsto -eQA_\mu\,,\quad r_\mu\mapsto -eQA_\mu\,,\quad t_{\mu\nu}\mapsto 0\,,
\end{align}
with $M=\mathrm{diag}(m_u,m_d,m_s)$ and $Q=\mathrm{diag}(2/3,-1/3,-1/3)$, while $\tilde{S}$, $\tilde{l}_\mu$, $\tilde{r}_\mu$, and $\tilde{t}_{\mu\nu}$ encode the contributions from the higher dimensional operators of LNEFT. Here we only concern terms that are linear in the LNEFT sources, since the Wilson coefficients contain a suppressed factor $1/\Lambda_\mathrm{EW}^2$ or $1/\Lambda_\mathrm{NP}^2$ that the higher order terms can be neglected safely~\cite{Dekens:2018pbu}.

As discussed in Ref.~\cite{Dekens:2018pbu}, the matrix element  via a scalar operator to a physical photon $\langle\gamma(p,\epsilon)|S|0\rangle$ vanishes due to Lorentz and gauge invariance, whereas the matrix element via a vector operator to a physical photon $\langle\gamma(p,\epsilon)|V^\mu|0\rangle$ also vanishes due to gauge invariance as well as the on-shell photon condition. Hence only tensor operator contributes to $\mu\to e\gamma$. The results can also be obtained from the observation that, at the leading order of $U$ matrix expansion ($U=1$), there is no $F_{\mu\nu}$ term in the scalar $\chi$PT Lagrangian, whereas only the term proportional to $\partial^\nu F_{\mu\nu}$ is in the vector $\chi$PT Lagrangian~\cite{Dekens:2018pbu}. This observation also holds for the case of the NMMs since, in the $\chi$PT the external fields are model independent that they can be endowed with any possible physical implication. Given our aim is to investigate the nonperturpative effects in the NMMs, that will only involve the dimension-5 dipole operators and dimension-6 tensor operators of LNEFT. Therefore, we only write down the matching from the LNEFT tensor operators onto the $\chi$PT operators with tensor external sources. For operators matching included the scalar and vector operators between the two theories, one is referred to Ref.~\cite{Dekens:2018pbu}. The matching up to chiral counting $\mathcal{O}(p^4)$ is given by~\cite{Dekens:2018pbu}
\begin{align}\label{eq:tensor}
\bar{q}_L\sigma^{\mu\nu}\tilde{t}_{\mu\nu}q_R\to\Lambda_1\langle\tilde{t}_{\mu\nu}(UF_L^{\mu\nu}+F_R^{\mu\nu}U)\rangle+i\Lambda_2\langle\tilde{t}_{\mu\nu}D_\mu UU^\dagger D_\nu U\rangle+\mathcal{O}(p^6)\,,
\end{align}
where $D_\mu U=\partial_\mu U-ir_\mu U+iUl_\mu$ denotes the covariant derivative for Goldstone bosons, $\Lambda_{1,2}$ are the tensor nonperturbative low-energy constants~\cite{Cata:2007ns,Dekens:2018pbu}, and $\langle\cdots\rangle$ stands for the trace in flavor space. The matching of $\bar{q}_R\sigma^{\mu\nu}\tilde{t}_{\mu\nu}^\dagger q_L$ is given by the Hermitian conjugate of Eq.~\eqref{eq:tensor}. Making use of Eq.~\eqref{eq:split}, the electromagnetic field are contained in the field strength tensors $F_{L,R}^{\mu\nu}$ of Eq.~\eqref{eq:tensor}, with $F_L^{\mu\nu}=\partial_\mu l_\nu-\partial_\nu l_\mu-i[l_\mu,l_\nu]$ and  $F_R^{\mu\nu}=\partial_\mu r_\nu-\partial_\nu r_\mu-i[r_\mu,r_\nu]$. Then the matching from LNEFT tensor operators onto $\chi$PT operators with tensor external sources is straightforward:
\begin{align}
(\bar \nu_{Lp} \sigma^{\mu\nu} N_{Rr})(\bar q_{L} \sigma_{\mu\nu} q_{R}) &\to -2eQ_q\Lambda_1(\bar \nu_{Lp} \sigma^{\mu\nu} N_{Rr})F_{\mu\nu}+\mathcal{O}(p^6)\,,\label{eq:nuNqq}\\[0.2cm]
(\bar \nu_{Lp}^c \sigma^{\mu\nu} \nu_{Lr})(\bar q_{R} \sigma_{\mu\nu} q_{L}) &\to -2eQ_q\Lambda_1(\bar \nu_{Lp}^c \sigma^{\mu\nu} \nu_{Lr})F_{\mu\nu}+\mathcal{O}(p^6)\,,\label{eq:nunuqq}\\[0.2cm]
(\bar N_{Rp}^c \sigma^{\mu\nu} N_{Rr})(\bar q_{L} \sigma_{\mu\nu} q_{R}) &\to -2eQ_q\Lambda_1(\bar N_{Rp}^c \sigma^{\mu\nu} N_{Rr})F_{\mu\nu}+\mathcal{O}(p^6)\,,\label{eq:NNqq}
\end{align}
where $F_{\mu\nu}=\partial_\mu A_\nu-\partial_\nu A_\mu$ denotes the photon field-strength tensor. Comparing terms in the right side of Eqs.~\eqref{eq:nuNqq}-\eqref{eq:NNqq} with the dipole operators listed in table~\ref{tab:LNC} and~\ref{tab:LNV}, one immediately obtains the following additional nonperturbative contributions to the dipole operators:
\begin{align}
\delta L_{\substack{\nu N\gamma\\pr}}&=e c_T\frac{F_\pi^2}{\Lambda_\chi}\left[\frac{2}{3}L_{\substack{\nu Nd\\prdd}}^{T,RR}+\frac{2}{3}L_{\substack{\nu Nd\\prss}}^{T,RR}-\frac{4}{3}L_{\substack{\nu Nu\\pruu}}^{T,RR}\right]\,,\label{eq:dLnuNgam}\\[0.2cm]
\delta L_{\substack{\nu\gamma\\pr}}&=e c_T\frac{F_\pi^2}{\Lambda_\chi}\left[\frac{2}{3}L_{\substack{\nu d\\prdd}}^{T,LL}+\frac{2}{3}L_{\substack{\nu d\\prss}}^{T,LL}-\frac{4}{3}L_{\substack{\nu u\\pruu}}^{T,LL}\right]\,,\label{eq:dLnugam}\\[0.2cm]
\delta L_{\substack{N\gamma\\pr}}&=e c_T\frac{F_\pi^2}{\Lambda_\chi}\left[\frac{2}{3}L_{\substack{Nd\\prdd}}^{T,RR}+\frac{2}{3}L_{\substack{Nd\\prss}}^{T,RR}-\frac{4}{3}L_{\substack{Nu\\pruu}}^{T,RR}\right]\,,\label{eq:dLNgam}
\end{align}
where in the right side of each equality we have used the formula 
\begin{align}
\Lambda_1\sim c_T\frac{F_\pi^2}{\Lambda_\chi}=c_T\frac{\Lambda_\chi}{16\pi^2}\,,
\end{align}
which is obtained with the naive dimensional analysis (NDA)~\cite{Manohar:1983md,Gavela:2016bzc}. For numerical input of the constant $c_T$, the model estimate of Ref.~\cite{Mateu:2007tr} gives $c_T\approx-3.2$, while using the lattice input~\cite{Baum:2011rm} combined with $\chi$PT developed in Ref.~\cite{Cata:2007ns} and the resonance chiral theory (R$\chi$T)~\cite{Ecker:1988te,Ecker:1989yg} yields $c_T\approx-1.0(2)$~\cite{Cata:2008zc,Miranda:2018cpf,Chen:2019vbr,Husek:2020fru}. Note that the two published estimates of $c_T$ are disagreed with each other by the amount $(3.2-1.0)/0.2=11\sigma$, this considerable uncertainty will affect the extractions of limits on the LNEFT coefficients. We will detail this aspect in next section. As stated in Sec.~\ref{sec:introduction}, both the Wilson coefficients $L_{iq}^{T,AA}$ and $c_T$ in Eqs.~\eqref{eq:dLnuNgam}-\eqref{eq:dLNgam} are evaluated at the renormalization scale $\mu=2~\mathrm{GeV}$. Note that, however, the $\mu$ dependence due to gluon corrections cancels in their product~\cite{Dekens:2018pbu}. 

\section{Constraints and discussions}\label{sec:numerical}

In this section, we shall constrain the Wilson coefficients of LNEFT dipole and tensor operators from two processes that the NMMs have contributions: the neutrino-electron scattering and the CE$\nu$NS. The right-handed neutrinos mass $m_{N_R}$ in LNEFT can be arbitrary, so one can always investigate the bounds for parameters that are of $m_{N_R}$ dependence. Nevertheless, since our aim in this paper is to study the nonperturbative effects of NMMs, a fixed value of $m_{N_R}$ will be sufficient to illustrate our purpose. Therefore, for simplicity we will restrict to the massless limit: $m_{N_R}\to0$. We also assume that the coefficients are real for simplicity, and work under the ``single-coefficient-dominance'' assumption for the purpose to obtain the maximum upper bounds on the LNEFT coefficients. Note, however, we need to stress that in a realistic UV complete model, there is no reason to believe that only a single operator is present at the experiment scale. Therefore, in general, when one uses experimental constraints to put limits on coefficients of operators in effective Lagrangians, the constraints are on a multidimensional space, since many different operators contribute and their coefficients are often complex.  Besides, we will also neglect the contributions from other LNEFT operators, which have been worked out in the literatures, see e.g. Refs.~\cite{Lindner:2016wff,AristizabalSierra:2018eqm,Giunti:2019xpr,Chang:2020jwl,Babu:2020ivd,Shoemaker:2020kji,Brdar:2020quo,Han:2020pff,Li:2020lba}. We will first present the constraints at the renormalization scale $\mu=2~\mathrm{GeV}$, the results then can be evolved onto the values at scale $\mu=\Lambda_\mathrm{EW}$ (which we choose $\Lambda_\mathrm{EW}=m_W$) by performing the RG equations. This procedure allows us to display the competition between the nonperturbative and perturbative effects in different processes. 

Given that the experiment scales where the Borexino and COHERENT take place are below the 2~GeV reference RG scale, to estimate nonperturbative effects at the reference scale, we need to run the experiment bounds from the lower energies up to 2~GeV. For CE$\nu$NS in COHERENT experiment, although the neutrino energies involved are in the range 20-50~MeV that is far below 2~GeV, the RG scale for this experiment can be chosen to be about 1~GeV, which corresponds to the mass of the nucleus in the effective field theory, below that scale the contributions to the matrix elements are described by the nucleon form factors, see Sec.~\ref{sec:CENUNSBounds} for more details. As both QCD and QED running effects for the relevant coefficients from 1~GeV to 2~GeV are very small, they can be neglected safely. It is noteworthy that with the same COHERENT data the bounds on the Wilson coefficients at the renormalization scales 1~GeV as well as 2~GeV have also been studied respectively in Refs.~\cite{Li:2020lba} and~\cite{Han:2020pff}. As for neutrino-electron scattering in Borexino, similarly, the the RG scale for experiment can be taken as the electron mass, 0.5~MeV, which is orders of magnitude smaller than 2~GeV. With the assumption that the scattering is triggered by the neutrino magnetic moments, which is described by the LNEFT dipole operators, we need to run the corresponding dipole coefficients from 0.5~MeV up to 2~GeV. Fortunately, as shown in Eqs.~\eqref{eq:LNCRGE} and~\eqref{eq:LNVRGE} the self-renormalization for the dipole coefficients is governed by the pure QED RG equations, 
\begin{equation}
\dot{L}_{\substack{i\gamma\\pr}}=-b_{0,e}e^2L_{\substack{i\gamma\\pr}}\,,\quad i=\nu,N,\nu N\,,
\end{equation}
and their exact solutions are
\begin{equation}
L_{\substack{i\gamma\\pr}}(\mu_2)=\left[\frac{\alpha(\mu_1)}{\alpha(\mu_2)}\right]^{\frac{-b_{0,e}}{2b_{0,e}}}L_{\substack{i\gamma\\pr}}(\mu_1)\,,\quad \mu_2<\mu_1\,.
\end{equation}
With $\mu_1=2$~GeV and $\mu_2=0.5$~MeV, we obtain
\begin{equation}
L_{\substack{i\gamma\\pr}}(0.5~\mathrm{MeV})\approx0.994L_{\substack{i\gamma\\pr}}(2~\mathrm{GeV})\,,
\end{equation}
which is the corollary of the small value as well as the small running effect of the QED coupling. This implies that the dipole coefficients at 2~GeV extremely close to the ones at experiment scales. Therefore, it is safe for us to take the values of the Wilson coefficients at the experiment scales as the ones at 2~GeV, which ensures us to reliably estimate the nonperturbative effects at 2~GeV.

\subsection{Neutrino-electron scattering}

Up to date, the most stringent bounds for NMMs on the terrestrial experiments are from the (anti-) neutrino-electron scattering. For instance, this can be applied to the short-baseline experiment GEMMA experiment with reactor antineutrinos~\cite{Beda:2012zz}, and to the long-baseline experiment Borexino experiment with solar neutrinos~\cite{Borexino:2017fbd}. The XENON1T excess may also be explained by the neutrino-electron scattering mediated by NMM~\cite{XENON:2020rca}, with neutrinos emitted from solar, for more dedicated studies in this regard see e.g. Refs.~\cite{Babu:2020ivd,Shoemaker:2020kji,Brdar:2020quo}. 

In the presence of additional NMM, the total differential cross section for neutrino-electron scattering can be written as~\cite{Vogel:1989iv}
\begin{align}\label{eq:dsigmamu}
\frac{d \sigma_{\nu e\mathrm{S}}}{d E_{r}}=\frac{d \sigma_{\mathrm{SM}}}{d E_{r}}+\alpha \mu_\nu^{2}\left[\frac{1}{E_{r}}-\frac{1}{E_{\nu}}\right]\,,
\end{align}
where $E_{r}$ denotes the recoil electron kinetic   energy, whereas $E_\nu$ is the incoming neutrino energy. In Eq.~\eqref{eq:dsigmamu}, $\mu_\nu$ is an effective magnetic moment, which accounts the informations of neutrino mixing and oscillations, and is a function of the propagation distance $L$ from the source to the scattering point and neutrino energy $E_\nu$~\cite{Borexino:2017fbd},
\begin{align}
\mu_\nu^2(L,E_\nu)=\sum_j\Big|\sum_k\mu_{jk}A_k(L,E_\nu)\Big|^2\,,
\end{align}
where $\mu_{jk}$ is an element of the NMMs matrix (in the neutrino mass eigenstates), and $A_k(L,E_\nu)$ is the amplitude of the $k$-mass state at the point of scattering~\cite{Borexino:2017fbd,Beacom:1999wx}. The initial neutrino of the scattering can be determined, but the final state of neutrino can be arbitrary due to its invisible property. Thus, Eq.~\eqref{eq:dsigmamu} allows us to constrain coefficients of different types of neutrino dipole operators. In Eq.~\eqref{eq:dsigmamu}, the expression for neutrino-electron scattering in the SM is given by~\cite{Bahcall:1995mm,Xing:2014wwa}:
\begin{align}\label{eq:SMnuescattering}
\frac{d \sigma_{\mathrm{SM}}}{d E_{r}}=\frac{G_F^2m_e}{2\pi E_\nu^2}\left[g_+^2E_\nu^2+g_-^2(E_\nu-E_r)^2-g_+g_-m_eE_r\right]\,,
\end{align}
where $G_F$ is the Fermi constant, $m_e$ is the electron mass, and
\begin{align}
\left\lbrace
\begin{array}{ll}
g_+=2\sin^2\theta_W+1 & \qquad\text{for}~\nu=\nu_e\\
g_+=2\sin^2\theta_W-1 & \qquad\text{for}~\nu=\nu_{\mu,\tau}\\
g_-=2\sin^2\theta_W   & \qquad\text{for}~\nu=\nu_{e,\mu,\tau}
\end{array}
\right.\,,
\end{align}
with $\theta_W$ being the Weinberg angle and numerically $\sin^2\theta_W=0.23857$~\cite{ParticleDataGroup:2020ssz}. The inputs of $g_+$ for $\nu_e$ and $\nu_{\mu,\tau}$ are different because, either charged or neutral weak current has the contribution to the $\nu_e$-$e$ scattering, while only neutral weak current can contribute to the $\nu_{\mu,\tau}$-$e$ scattering. A similar expression for the antineutrino-electron scattering can be obtained by simply exchanging the positions of $g_+$ and $g_-$ in Eq.~\eqref{eq:SMnuescattering}~\cite{Xing:2014wwa}.

We focus on the upper constraints from the terrestrial experiments. Using the solar neutrinos as the sources, the Borexino experiment gives the most stringent upper bound on the NMM, which at the $90\%$ confidence level (CL) reads~\cite{Borexino:2017fbd}
\begin{align}\label{eq:Borexinobound}
\mu_\nu<2.8\times10^{-11}\mu_B\,.
\end{align} 
Similarly, another terrestrial experiment, GEMMA, which uses the reactor antineutrinos as sources, obtains a slightly weaker bound at the $90\%$ CL~\cite{Beda:2012zz},
\begin{align}
\mu_\nu<2.9\times10^{-11}\mu_B\,.
\end{align}

Here, we shall adopt the result of Eq.~\eqref{eq:Borexinobound} in our numerical analysis.
For the purpose to simplify the numerical computations, we assume that the Dirac NMMs are flavor diagonal and universal, then the dependence on the oscillation parameters cancels out. Similarly, we also assume that all the Majorana transition magnetic moments carry the same value. Therefore, the effective magnetic moment $\mu_\nu$ measured from Borexino can connect to the Wilson coefficients of LNEFT dipole operators via the following relation:
\begin{align}\label{eq:muLNEFT}
\mu_\nu^2=\left\lbrace
\begin{array}{ll}
\Big|2L_{\substack{i\gamma\\pr}}\Big|^2 & \qquad\text{for}~i=\nu N\\
\Big|4L_{\substack{i\gamma\\pr}}\Big|^2 & \qquad\text{for}~i=\nu
\end{array}
\right.\,.
\end{align}
The limits for the LNEFT coefficients then can be read off directly after combining Eqs.~\eqref{eq:Borexinobound} and~\eqref{eq:muLNEFT}. Using the one-operator-at-a-time constraint, we collect in tables~\ref{tab:nueSLNC} and~\ref{tab:nueSLNV} the upper bounds for the Wilson coefficients of the LNC and LNV LNEFT dipole and tensor operators at the renormalization scales $\mu=2~\mathrm{GeV}$ as well as $\mu=m_W$. The bounds are given in units of $\mathrm{GeV}^{-1}$ for the dipole coefficients ($d=5$), while those for tensor coefficients ($d=6$) are in units of $\mathrm{GeV}^{-2}$. The tensor operators can contribute to neutrino-electron scattering through two ways: on the one hand, the tensor operators can contribute to the  dipole operators through the nonperturbative
matching effects which are proportional to $c_T$, and on the other hand, the RG evolutions from $\mu=m_W$ down to $\mu=2~\mathrm{GeV}$ induce perturbative mixing contributions to the coefficients of the dipole operators which are independent of $c_T$.
Clearly, as the magnitude of $c_T$ is of order $\mathcal{O}(1)$, the numerical results in the third rows of tables~\ref{tab:nueSLNC} and~\ref{tab:nueSLNV} imply that the nonperturbative contributions dominate for the LNEFT Wilson coefficients involving up and down quarks, while they are expected to be of the same order of magnitude as the perturbative contributions for the couplings to strange quark. Note that similar conclusions have been obtained in Ref.~\cite{Dekens:2018pbu} for the charged lepton-quark interactions.
\begin{table}[!thb]
\centering
\renewcommand{\arraystretch}{1.75}
\begin{tabular}{|c|c|c|c|c|}
\hline
$\mu=2~\mathrm{GeV}$ & $\Big|L_{\substack{\nu N\gamma\\\ell\ell}}\Big|$ & $\Big|c_TL^{T,RR}_{\substack{\nu Nu\\\ell\ell uu}}\Big|$ & $\Big|c_TL^{T,RR}_{\substack{\nu Nd\\\ell\ell dd}}\Big|$ & $\Big|c_TL^{T,RR}_{\substack{\nu Nd\\\ell\ell ss}}\Big|$\\[0.2cm]
\hline 
$[\mathrm{GeV}^{4-d}]$ & $4.2\times10^{-9}$ & $1.4\times10^{-6}$ & $2.8\times10^{-6}$ & $2.8\times10^{-6}$\\
\hline
$\mu=m_W$ & $\Big|L_{\substack{\nu N\gamma\\\ell\ell}}\Big|$ & $\Big|(c_T-0.08)L^{T,RR}_{\substack{\nu Nu\\\ell\ell uu}}\Big|$ & $\Big|(c_T-0.18)L^{T,RR}_{\substack{\nu Nd\\\ell\ell dd}}\Big|$ & $\Big|(c_T-3.55)L^{T,RR}_{\substack{\nu Nd\\\ell\ell ss}}\Big|$\\[0.2cm]
\hline 
$[\mathrm{GeV}^{4-d}]$ & $4.3\times10^{-9}$ & $1.6\times10^{-6}$ & $3.3\times10^{-6}$ & $3.3\times10^{-6}$\\
\hline
\end{tabular}
\caption{Upper bounds for Wilson coefficients of the LNC LNEFT dipole and tensor operators (with $\ell=e,\mu,\tau$) obtained from neutrino-electron scattering in Borexino~\cite{Borexino:2017fbd}. The results in the second (fourth) row for the corresponding coefficients in first (third) row are obtained at the renormalization scale $\mu=2~\mathrm{GeV}$ ($\mu=m_W$). In the third row, the parameter $c_T$ in the parentheses indicates the nonperturbative contributions, while the remaining values represent the relative perturbative contributions.}\label{tab:nueSLNC}
\end{table} 

\begin{table}[!thb]
\centering
\renewcommand{\arraystretch}{1.75}
\begin{tabular}{|c|c|c|c|c|}
\hline
$\mu=2~\mathrm{GeV}$ & $\Big|L_{\substack{\nu \gamma\\pr}}\Big|$ & $\Big|c_TL^{T,LL}_{\substack{\nu u\\pr uu}}\Big|$ & $\Big|c_TL^{T,LL}_{\substack{\nu d\\pr dd}}\Big|$ & $\Big|c_TL^{T,LL}_{\substack{\nu d\\pr ss}}\Big|$\\[0.2cm]
\hline 
$[\mathrm{GeV}^{4-d}]$ & $2.1\times10^{-9}$ & $0.7\times10^{-6}$ & $1.4\times10^{-6}$ & $1.4\times10^{-6}$\\
\hline
$\mu=m_W$ & $\Big|L_{\substack{\nu \gamma\\pr}}\Big|$ & $\Big|(c_T-0.08)L^{T,LL}_{\substack{\nu u\\pr uu}}\Big|$ & $\Big|(c_T-0.18)L^{T,LL}_{\substack{\nu d\\pr dd}}\Big|$ & $\Big|(c_T-3.55)L^{T,LL}_{\substack{\nu d\\pr ss}}\Big|$\\[0.2cm]
\hline 
$[\mathrm{GeV}^{4-d}]$ & $2.2\times10^{-9}$ & $0.8\times10^{-6}$ & $1.7\times10^{-6}$ & $1.7\times10^{-6}$\\
\hline
\end{tabular}
\caption{The same for Wilson coefficients of the LNV LNEFT dipole and tensor operators. The dipole and tensor operators are antisymmetric in the flavor indices, so only those with $p\neq r$ are nonvanishing.}\label{tab:nueSLNV}
\end{table}

\subsection{Coherent elastic neutrino-nucleus scattering}
\label{sec:CENUNSBounds}

In the CE$\nu$NS, the low-energy neutrinos can couple to protons and neutrons of the nucleus coherently, so its cross section can be significantly enhanced. This implies that the CE$\nu$NS can not only provide a precision test of neutrino interactions in the SM, but also put very stringent bounds on parameters from NP.

In the presence of the additional dipole operators and tensor operators, the differential cross section for CE$\nu$NS can be written as~\cite{Lindner:2016wff,Chang:2020jwl,Li:2020lba}
\begin{align}\label{eq:CEnuNS}
\frac{d\sigma_\mathrm{CE\nu NS}}{dE_r}=\frac{G_F^2M}{4\pi}\left[\xi^2_V\left(1-\frac{E_r}{E_r^\mathrm{max}}\right)+\xi^2_T\left(1-\frac{E_r}{2E_r^\mathrm{max}}\right)+e^2A_M^2(\frac{1}{ME_r}-\frac{1}{ME_\nu})\right]\,,
\end{align}
where $M$ is the mass of nucleus, $E_\nu$ denotes the incoming neutrino energy, and $E_r$ stands for the recoil nucleus kinetic energy with maximal value $E_r^\mathrm{max}=\frac{2E_\nu^2}{M+2E_\nu}\approx\frac{2E_\nu^2}{M}$. The SM contribution is contained in parameter $\xi_V^2$, with
\begin{align}
\xi_{V,\mathrm{SM}}^2=\left[\mathbb{N}-(1-4\sin^2\theta_W)\mathbb{Z}\right]^2F^2(q^2)\,,
\end{align}
where $\mathbb{Z}$ and $\mathbb{N}$ denote respectively the proton and neutron numbers of a given nucleus $\mathcal{N}$, and $F(q^2)$ is the Helm form factor of the nucleus (with $q$ being the transferred energy) whose coherent limit $(q^2\to0)$ is 1~\cite{Helm:1956zz}.\footnote{Here we have assumed that all the proton and neutron form factors are equal to the Helm form factor, i.e. $F_p(q^2)=F_n(q^2)=F(q^2)$.} In comparison, the dipole and tensor contributions from NP are encoded in $A_M^2$ and $\xi_T^2$, respectively. Clearly, there is no interference between the SM term and other NP contributions. Similar to the case of neutrino-electron scattering, the incoming neutrino of CE$\nu$NS can be controlled, but the neutrino in the final state can be either left-handed or right-handed. Therefore, this allows us to constrain coefficient of either LNC or LNV operators, concretely,
\begin{align}
A_M^2
&=\left\lbrace
\begin{array}{ll}
\sum\limits_{r}\big|\frac{2}{G_F}L_{\substack{i\gamma\\pr}}\big|^2\mathbb{Z}^2F^2(q^2) & \text{for}~i=\nu N\\[0.2cm]
\sum\limits_{r}\big|\frac{4}{G_F}L_{\substack{i\gamma\\pr}}\big|^2\mathbb{Z}^2F^2(q^2) & \text{for}~i=\nu
\end{array}
\right.\,,\label{eq:AM}\\[0.2cm]
\xi_T^2
&=\left\lbrace
\begin{array}{ll}
8\sum\limits_{r,j}\Big|\frac{\sqrt{2}}{G_F}\sum\limits_{q=u,d,s}L_{\substack{iu/d\\prqq}}^{T,RR}\left(\mathbb{Z}_j\delta_q^p+\mathbb{N}_j\delta_q^n\right)\Big|^2F^2(q^2) & \text{for}~i=\nu N\\[0.2cm]
8\sum\limits_{r,j}\Big|\frac{2\sqrt{2}}{G_F}\sum\limits_{q=u,d,s}L_{\substack{iu/d\\prqq}}^{T,LL}\left(\mathbb{Z}_j\delta_q^p+\mathbb{N}_j\delta_q^n\right)\Big|^2F^2(q^2) & \text{for}~i=\nu
\end{array}
\right.\,,\label{eq:xiT}
\end{align}
where the subscript $j$ sums over the nucleus that participates the scattering. For example, the COHERENT experiment uses CsI as the detector which be exposed to the neutrino emissions ($\nu_\mu$, $\bar{\nu}_\mu$, and $\nu_e$) from the Spallation Neutron Source at Oak Ridge National Laboratory~\cite{COHERENT:2017ipa}, from which the proton and neutron numbers for Caesium and Iodine are $\mathbb{Z}_\mathrm{Cs}=55$, $\mathbb{N}_\mathrm{Cs}=77.9$, $\mathbb{Z}_\mathrm{I}=53$, and $\mathbb{N}_\mathrm{I}=73.9$, respectively. The functions $\delta_q^{p/n}$ in Eq.~\eqref{eq:xiT} are the nucleon form factors for tensor current, with numerical values~\cite{Belanger:2008sj,Belanger:2018ccd}
\begin{align}\label{eq:nucleonFF}
\delta_u^p=0.84\,,\quad\delta_d^p=-0.23\,, \quad \delta_s^p=-0.046\,,\quad \delta_u^n=-0.23\,, \quad\delta_d^n=0.84\,, \quad \delta_s^n=-0.046\,.
\end{align}

By using the COHERENT data~\cite{COHERENT:2017ipa}, the $90\%$ CL bounds for dipole and tensor parameters in the one-operator-at-a-time constraint are given respectively by~\cite{AristizabalSierra:2018eqm,Chang:2020jwl}
\begin{align}\label{eq:CEnuNSbounds}
\frac{1}{2}\frac{A_M^2}{v^2\mathbb{Z}^2F^2(q^2)}<7.2\times10^{-8}\,,\qquad 
\frac{\xi^2_T}{\mathbb{N}^2F^2(q^2)}<0.591^2\,,
\end{align}
where $v=246~\mathrm{GeV}$ is vacuum expectation value of Higgs field. Combining Eqs.~\eqref{eq:AM}-\eqref{eq:CEnuNSbounds}, one immediately obtains constraints for the LNEFT Wilson coefficients. Note that from Eq.~\eqref{eq:CEnuNS} the tensor operator can contribute to CE$\nu$NS through both direct and indirect ways. The latter is similar to the case in neutrino-electron scattering, i.e., the tensor operators can contribute to the CE$\nu$NS either through the nonperturbative matching onto the dipole operators, which are proportional to $c_T$, or through the perturbative mixing to the dipole operators when taking into account the the RG evolutions from $\mu=m_W$ down to $\mu=2~\mathrm{GeV}$, which are independent of $c_T$. However, due to the smallness of the anomalous dimensions and the nonperturbative matching parameters, the process will be dominated by the direct contributions. As shown in tables~\ref{tab:CEnuNSLNC} and~\ref{tab:CEnuNSLNV}, the indirect contributions including nonperturbative and perturbative parts for couplings to up and down quarks become negligible in comparison with the direct ones. As for the coupling to strange quark, the nonperturbative contribution is the same order of magnitude as the perturbative contribution, and is about $15\%$ ($5\%$) of the direct contribution, using $c_T=-3.2$ ($c_T=-1.0(2)$). This can be attributed to two reasons: on the one hand, the larger anomalous dimension with strange quark mass can enhance the mixing effect, and on the other hand, the smaller nucleon form factors for the strange quark (see Eq.~\eqref{eq:nucleonFF}) can reduce the proportion of direct contribution. Note that in tables~\ref{tab:CEnuNSLNC} and~\ref{tab:CEnuNSLNV} the numerical results for LNEFT coefficients at $\mu=2~\mathrm{GeV}$ have also been presented in Ref.~\cite{Li:2020lba}.
\begin{table}[tbh!]
\begin{center}
\renewcommand{\arraystretch}{1.75}
\resizebox{\textwidth}{!}{
\begin{tabular}{|c|c|c|c|c|}
\hline
\multirow{2}{*}{$\mu=2~\mathrm{GeV}$}
& $\Big|L_{\substack{\nu N\gamma\\pr}}\Big|$ & $\Big|L^{T,RR}_{\substack{\nu Nu\\pr uu}}\Big|$ & $\Big|L^{T,RR}_{\substack{\nu Nd\\pr dd}}\Big|$ & $\Big|L^{T,RR}_{\substack{\nu Nd\\pr ss}}\Big|$\\
& & $\Big|0.018c_TL^{T,RR}_{\substack{\nu Nu\\pr uu}}\Big|$ & $\Big|0.005c_TL^{T,RR}_{\substack{\nu Nd\\pr dd}}\Big|$ & $\Big|0.041c_TL^{T,RR}_{\substack{\nu Nd\\pr ss}}\Big|$\\[0.2cm]
\hline 
$[\mathrm{GeV}^{4-d}]$ & $5.4\times10^{-7}$ & $3.3\times10^{-6}$ & $1.8\times10^{-6}$ & $1.5\times10^{-5}$\\
\hline
\multirow{2}{*}{\centering $\mu=m_W$}
& $\Big|L_{\substack{\nu N\gamma\\pr}}\Big|$ & $\Big|L^{T,RR}_{\substack{\nu Nu\\pr uu}}\Big|$ & $\Big|L^{T,RR}_{\substack{\nu Nd\\pr dd}}\Big|$ & $\Big|L^{T,RR}_{\substack{\nu Nd\\pr ss}}\Big|$\\
& & $\Big|(0.021c_T-0.0017)L^{T,RR}_{\substack{\nu Nu\\pr uu}}\Big|$ & $\Big|(0.006c_T-0.001)L^{T,RR}_{\substack{\nu Nd\\pr dd}}\Big|$ & $\Big|(0.048c_T-0.17)L^{T,RR}_{\substack{\nu Nd\\pr ss}}\Big|$\\[0.2cm]
\hline 
$[\mathrm{GeV}^{4-d}]$ & $5.5\times10^{-7}$ & $3.9\times10^{-6}$ & $2.1\times10^{-6}$ & $1.8\times10^{-5}$\\
\hline
\end{tabular}}
\caption{Upper bounds for Wilson coefficients of the LNC LNEFT dipole and tensor operators (with $p=e,\mu$) obtained from CE$\nu$NS in COHERENT~\cite{COHERENT:2017ipa}. The results in second and fourth rows correspond to bounds from direct contributions. For ease of comparison, the indirect terms (including nonperturbative and perturbative contributions), which are obtained by normalizing the dipole bounds to the respective tensor bounds, have been written below the corresponding direct terms in first and third rows.}\label{tab:CEnuNSLNC}
\end{center}
\end{table} 
\begin{table}[tbh!]
\centering
\renewcommand{\arraystretch}{1.75}
\resizebox{\textwidth}{!}{
\begin{tabular}{|c|c|c|c|c|}
\hline
\multirow{2}{*}{\centering $\mu=2~\mathrm{GeV}$}
& $\Big|L_{\substack{\nu \gamma\\pr}}\Big|$ & $\Big|L^{T,LL}_{\substack{\nu u\\pr uu}}\Big|$ & $\Big|L^{T,LL}_{\substack{\nu d\\pr dd}}\Big|$ & $\Big|L^{T,LL}_{\substack{\nu d\\pr ss}}\Big|$\\
& & $\Big|0.018c_TL^{T,LL}_{\substack{\nu u\\pr uu}}\Big|$ & $\Big|0.005c_TL^{T,LL}_{\substack{\nu d\\pr dd}}\Big|$ & $\Big|0.041c_TL^{T,LL}_{\substack{\nu d\\pr ss}}\Big|$\\[0.2cm]
\hline 
$[\mathrm{GeV}^{4-d}]$ & $2.7\times10^{-7}$ & $1.7\times10^{-6}$ & $0.9\times10^{-6}$ & $7.7\times10^{-6}$\\
\hline
\multirow{2}{*}{\centering $\mu=m_W$}
& $\Big|L_{\substack{\nu\gamma\\pr}}\Big|$ & $\Big|L^{T,LL}_{\substack{\nu u\\pr uu}}\Big|$ & $\Big|L^{T,LL}_{\substack{\nu d\\pr dd}}\Big|$ & $\Big|L^{T,LL}_{\substack{\nu d\\pr ss}}\Big|$\\
& & $\Big|(0.021c_T-0.0017)L^{T,LL}_{\substack{\nu u\\pr uu}}\Big|$ & $\Big|(0.006c_T-0.001)L^{T,LL}_{\substack{\nu d\\pr dd}}\Big|$ & $\Big|(0.048c_T-0.17)L^{T,LL}_{\substack{\nu d\\pr ss}}\Big|$\\[0.2cm]
\hline 
$[\mathrm{GeV}^{4-d}]$ & $2.8\times10^{-7}$ & $2.0\times10^{-6}$ & $1.1\times10^{-6}$ & $9.0\times10^{-6}$\\
\hline
\end{tabular}}
\caption{The same for Wilson coefficients of the LNV LNEFT dipole and tensor operators.}\label{tab:CEnuNSLNV}
\end{table}

\subsection{Discussions}

As mentioned above, since the nonperturbative contributions crucially depend on the input of $c_T$, we can exactly calculate the upper limits of the tensor Wilson coefficients provided that the experimental bounds as well as the value of $c_T$ are known. Note that, however, the model and lattice estimates of $c_T$ quoted are strongly disagreed to each other, this will affect the extraction of limits on the coefficients. Combining the numerical results in tables~\ref{tab:nueSLNC}-\ref{tab:CEnuNSLNV} separately with the inputs of $c_T$, i.e., the model estimate $c_T=-3.2$ and the lattice estimate $c_T=-1.0(2)$, we list in tables~\ref{tab:TensorWCLNC} and~\ref{tab:TensorWCLNV} the upper bounds on the Wilson coefficients of the LNC and LNV LNEFT tensor operators at $2~\mathrm{GeV}$, respectively. Comparing the constraints from Borexino and COHERENT experiments, it is found that when using $c_T=-3.2$ as input, all bounds for LNEFT Wilson coefficients obtained from neutrino-electron scattering in Borexino are more stringent than the ones obtained from CE$\nu$NS in COHERENT. Nevertheless, when using $c_T=-1.0(2)$ as input, the above observation only holds true for the coefficients of the dipole operators as well as the tensor operators involving up and strange quarks, but it reverses for the tensor coefficients involving down quark. Following steps discussed above, the similar constraints and so the inferences at $\mu=m_W$ can also be acquired directly, so we will not discuss them any further.
\begin{table}[tbh!]
\centering
\renewcommand{\arraystretch}{1.75}
\setlength{\tabcolsep}{6mm}{
\begin{tabular}{|c|c|c|c|c|}
\hline
$\mu=2~\mathrm{GeV}$ & $\Big|L_{\substack{\nu N\gamma\\pr}}\Big|$ & $\Big|L^{T,RR}_{\substack{\nu N u\\pr uu}}\Big|$ & $\Big|L^{T,RR}_{\substack{\nu Nd\\pr dd}}\Big|$ & $\Big|L^{T,RR}_{\substack{\nu Nd\\pr ss}}\Big|$\\[0.2cm]
\hline 
\makecell[c]{Borexino\\$[\mathrm{GeV}^{4-d}]$} & $4.2\times10^{-9}$ & \makecell[c]{$4.4\times10^{-7}$\\$(1.7\times10^{-6})$} & \makecell[c]{$8.8\times10^{-7}$\\$(3.4\times10^{-6})$} & \makecell[c]{$8.8\times10^{-7}$\\$(3.4\times10^{-6})$}\\[0.2cm]
\hline 
\makecell[c]{COHERENT\\$[\mathrm{GeV}^{4-d}]$} & $5.4\times10^{-7}$ & $3.3\times10^{-6}$ & $1.8\times10^{-6}$ & $1.5\times10^{-5}$\\
\hline
\end{tabular}}
\caption{The upper bounds on the Wilson coefficients of the LNC LNEFT tensor operators at $2~\mathrm{GeV}$. For constraints from Borexino, the results outside (inside) the parentheses correspond to the limits with input $c_T=-3.2$ ($c_T=-1.0(2)$). In COHERENT constraints, we only consider direct contributions since, as discussed in Sec.~\ref{sec:CENUNSBounds}, the indirect contributions are negligible in comparison with the direct ones.}\label{tab:TensorWCLNC}
\end{table}

\begin{table}[tbh!]
\centering
\renewcommand{\arraystretch}{1.75}
\setlength{\tabcolsep}{6mm}{
\begin{tabular}{|c|c|c|c|c|}
\hline
$\mu=2~\mathrm{GeV}$ & $\Big|L_{\substack{\nu\gamma\\pr}}\Big|$ & $\Big|L^{T,LL}_{\substack{\nu  u\\pr uu}}\Big|$ & $\Big|L^{T,LL}_{\substack{\nu d\\pr dd}}\Big|$ & $\Big|L^{T,LL}_{\substack{\nu d\\pr ss}}\Big|$\\[0.2cm]
\hline 
\makecell[c]{Borexino\\$[\mathrm{GeV}^{4-d}]$} & $2.1\times10^{-9}$ & \makecell[c]{$2.2\times10^{-7}$\\$(0.8\times10^{-6})$} & \makecell[c]{$4.4\times10^{-7}$\\$(1.7\times10^{-6})$} & \makecell[c]{$4.4\times10^{-7}$\\$(1.7\times10^{-6})$}\\[0.2cm]
\hline 
\makecell[c]{COHERENT\\$[\mathrm{GeV}^{4-d}]$} & $2.7\times10^{-7}$ & $1.7\times10^{-6}$ & $0.9\times10^{-6}$ & $7.7\times10^{-6}$\\
\hline
\end{tabular}}
\caption{The same for the upper bounds on the Wilson coefficients of the LNV LNEFT tensor operators at $2~\mathrm{GeV}$.}\label{tab:TensorWCLNV}
\end{table}

The bounds on LNV dipole and tensor LNEFT coefficients with right-handed neutrinos can be studied in the vector meson decays (e.g. $\omega\to\mathrm{inv.}$ and $\phi\to\mathrm{inv.}$), from which the direct contributions will dominate for all couplings. Nevertheless, since the constraints from the current experimental data~\cite{ParticleDataGroup:2020ssz} are rather weak~\cite{Li:2020lba}, we do not include them in our paper.  

Generally speaking, the LNEFT Wilson coefficients can be complex, albeit throughout this paper we have impliedly assumed that $CP$ is conserved so that they are real numbers. This assumption is unnecessary, since our method can also be applied to the neutrino electric dipole moments (EDMs), which are proportional the imaginary part of  the LNEFT coefficients, provided that $CP$ is violated. For instance, 
if $CP$ is violated in the neutrino radiative decays, an asymmetry can be observed in the circularly polarised photons~\cite{Boehm:2017nrl,Balaji:2019fxd,Balaji:2020oig}.

The NMMs can also be indirectly estimated from the cosmological measurements as well as the astrophysical observations. The plasmon decays into neutrino pairs via nonzero NMM will lead to increased energy loss in stellar environments, thus the relevant astrophysical observations can provide an indirect constraint on the NMM. For instance, the observation from the red giant branch of globular clusters in this argument results in an upper limit on the NMM $4.5\times10^{-12}\mu_B$~\cite{Viaux:2013lha}, also, the observed neutrino signal from SN1987A (SN refers to supernova) leads to the upper bound a few $\times10^{-12}\mu_B$~\cite{Kamiokande-II:1987idp,Bionta:1987qt,Alekseev:1988gp,Magill:2018jla}, both are one order of magnitude more stringent than the one obtain from Borexino~\cite{Borexino:2017fbd}. As neutrinos are a dominant ingredient of the early Universe during the Big Bang Nucleosynthesis (BBN) era, this environment is very sensitive to the additional interactions triggered by the NMM. For Dirac NMM, with a right-handed neutrino decoupling temperature of $T_\mathrm{dec}\sim100~\mathrm{MeV}$ the upper limit is estimated to be $7\times10^{-11}\mu_B$ in Ref.~\cite{Fukugita:1987uy}, $6.2\times10^{-11}\mu_B$ in Ref.~\cite{Elmfors:1997tt}, and $2.9\times10^{-10}\mu_B$ in Ref.~\cite{Ayala:1999xn}, while for Majorana neutrino transition magnetic moment, the upper limit is of order $10^{-10}\mu_B$~\cite{Vassh:2015yza}, all these results are less severe than the one obtained from Borexino~\cite{Borexino:2017fbd}. In this work, however, we only focus on the bounds from the direct measurements of the terrestrial experiments, and exclude the indirect estimates from the cosmological measurements as well as astrophysical observations, for that a vast of regions of parameter space cannot be accommodated with the terrestrial experiments. One can account for the bounds from terrestrial experiments while evading the cosmological and astrophysical bounds by adding some extra fields, see e.g. the recent comprehensive analysis in this regard in Ref.~\cite{Brdar:2020quo}. However, they have been out of scope for this work, and we will not discuss them any further.

\section{Conclusions}\label{sec:conclusions}

In this paper, we have investigated the nonperturbative contributions of both the LNC and LNV LNEFT neutrino-quark tensor operators to the NMMs. The nonperturbative effects can be obtained by matching the LNEFT tensor operators onto the $\chi$PT operators with tensor external sources. The nonperturbative contributions are of order $L_{i\gamma}\sim e(F_\pi^2/\Lambda_\chi)L_{iq}^{T,AA}$, which can be compared to the perturbative contributions that are of order $L_{i\gamma}\sim e m_q/(16\pi^2)L_{iq}^{T,AA}$, and are obtained from one-loop mixing by performing the RG evolutions from $\mu=m_W$ down to $\mu=2~\mathrm{GeV}$. In order to show the competition between these two effects, we have constrained the relevant Wilson coefficients of LNC and LNV LNEFT dipole and tensor operators from the neutrino-electron scattering with the Borexino data~\cite{Borexino:2017fbd} and CE$\nu$NS with the COHERENT data~\cite{COHERENT:2017ipa}, the numerical results are shown in tables~\ref{tab:nueSLNC}-\ref{tab:CEnuNSLNV}. In the neutrino-electron scattering, it is found that the nonperturbative contributions dominate for the LNEFT Wilson coefficients involving up and down quarks, while they are expected to be of the same order of magnitude as the perturbative contributions for the couplings to strange quark. As for constraints in the CE$\nu$NS, the tensor operators can contribute to the process through either direct or indirect way. As a result, the indirect contributions including nonperturbative and perturbative parts for couplings to up and down quarks become negligible in comparison with the direct ones. As the nonperturbative contributions crucially depend on the value of $c_T$, its inputs affect the extraction of limits on the tensor LNEFT Wilson coefficients. We have calculate the upper bounds on these coefficients separately with the model estimate $c_T=-3.2$ and the lattice input $c_T=-1.0(2)$ at 2~GeV, the results are listed in tables~7 and~8. It is found that all bounds for LNEFT Wilson coefficients obtained from neutrino-electron scattering in Borexino are more stringent than the ones obtained from CE$\nu$NS in COHERENT, using $c_T=-3.2$. Nevertheless, when using $c_T=-1.0(2)$ as input, the above observation only holds true for the coefficients of the dipole operators as well as the tensor operators involving up and strange quarks, but it reverses for the tensor coefficients involving down quark.

\section*{Acknowledgements}

This work is supported in part by the National Natural Science Foundation of China under Grant No. 11875327, the Fundamental Research Funds for the Central Universities, and the Sun Yat-Sen University Science Foundation. F.C. is also supported by the CCNU-QLPL Innovation Fund (QLPL2021P01).

\appendix

\section{RG equations for the Wilson coefficients of LNEFT operators}\label{app:RGE}

The RG equations for a subset of Wilson coefficients of LNEFT operators have been computed in Refs.~\cite{Jenkins:2017dyc,Han:2020pff,Li:2020lba}, the anomalous dimension matrix of which can be obtained from one-loop QCD and QED corrections. Here we only list the RG equations for Wilson coefficients of both LNC and LNV dimension-5 dipole operators as well as dimension-6 neutrino-quark interacting operators that we will use throughout this paper. The RG equations for the LNC LNEFT Wilson coefficients are
\begin{align}\label{eq:LNCRGE}
\dot{L}_{\substack{\nu N\gamma\\pr}}&=-b_{0,e}e^2L_{\substack{\nu N\gamma\\pr}}-16e[M_u]_{wt}
L_{\substack{\nu Nu\\prwt}}^{T,RR}+8e[M_d]_{wt}L_{\substack{\nu Nd\\prwt}}^{T,RR}\,,\nonumber\\[0.2cm]
\dot{L}_{\substack{\nu Nu\\prwt}}^{T,RR}
&=\left[2g_s^2C_F+\frac{8}{9} e^2\right]L_{\substack{\nu Nu\\prwt}}^{T,RR}\,, 
\hspace{3cm} \dot{L}_{\substack{\nu Nd\\prwt}}^{T,RR}=\left[2g_s^2C_F+\frac{2}{9} e^2\right]L_{\substack{\nu Nd\\prwt}}^{T,RR}\,,\nonumber\\[0.2cm]
\dot{L}_{\substack{\nu Nu\\prwt}}^{S,RR}&=\left[-6g_s^2C_F-\frac{24}{9} e^2\right]L_{\substack{\nu Nu\\prwt}}^{S,RR}\,, 
\hspace{2.5cm}\dot{L}_{\substack{\nu Nd\\prwt}}^{S,RR}=\left[-6g_s^2C_F-\frac{6}{9} e^2\right]L_{\substack{\nu Nd\\prwt}}^{S,RR}\,,\nonumber\\[0.2cm]
\dot{L}_{\substack{\nu Nu\\prwt}}^{S,RL}&=\left[-6g_s^2C_F-\frac{24}{9} e^2\right]L_{\substack{\nu Nu\\prwt}}^{S,RL}\,,
\hspace{2.5cm}\dot{L}_{\substack{\nu Nd\\prwt}}^{S,RL}=\left[-6g_s^2C_F-\frac{6}{9} e^2\right]L_{\substack{\nu Nd\\prwt}}^{S,RL}\,,
\end{align}
where $C_F=\frac{4}{3}$, and $b_{0,e}=-\frac{4}{3}(n_e+\frac{1}{3}n_d+\frac{4}{3}n_u)$ is the leading coefficient of the QED beta function with $n_e$, $n_d$, and $n_u$ standing for the numbers of active charged lepton, down-type quarks, and up-type quarks, respectively. Similarly, the RG equations for the LNV LNEFT Wilson coefficients are
\begin{align}\label{eq:LNVRGE}
\dot{L}_{\substack{\nu\gamma\\pr}}&=-b_{0,e}e^2L_{\substack{\nu\gamma\\pr}}-16e[M_u]_{wt}
L_{\substack{\nu u\\prwt}}^{T,LL}+8e[M_d]_{wt}L_{\substack{\nu d\\prwt}}^{T,LL}\,,\nonumber\\[0.2cm]
\dot{L}_{\substack{N\gamma\\pr}}&=-b_{0,e}e^2L_{\substack{N\gamma\\pr}}-16e[M_u]_{wt}
L_{\substack{N u\\prwt}}^{T,RR}+8e[M_d]_{wt}L_{\substack{N d\\prwt}}^{T,RR}\,,\nonumber\\[0.2cm]
\dot{L}_{\substack{\nu u\\prwt}}^{T,LL}
&=\left[2g_s^2C_F+\frac{8}{9} e^2\right]L_{\substack{\nu u\\prwt}}^{T,LL}\,,
\hspace{3cm} \dot{L}_{\substack{\nu d\\prwt}}^{T,LL}=\left[2g_s^2C_F+\frac{2}{9} e^2\right]L_{\substack{\nu d\\prwt}}^{T,LL}\,,\nonumber\\[0.2cm]
\dot{L}_{\substack{\nu u\\prwt}}^{S,LL}
&=\left[-6g_s^2C_F-\frac{24}{9} e^2\right]L_{\substack{\nu u\\prwt}}^{S,LL}\,,
\hspace{2.5cm} \dot{L}_{\substack{\nu d\\prwt}}^{S,LL}=\left[-6g_s^2C_F-\frac{6}{9} e^2\right]L_{\substack{\nu d\\prwt}}^{S,LL}\,,
\nonumber\\[0.2cm]
\dot{L}_{\substack{N u\\prwt}}^{T,RR}
&=\left[2g_s^2C_F+\frac{8}{9} e^2\right]L_{\substack{N u\\prwt}}^{T,RR}\,,
\hspace{3cm} \dot{L}_{\substack{N d\\prwt}}^{T,RR}=\left[2g_s^2C_F+\frac{2}{9} e^2\right]L_{\substack{N d\\prwt}}^{T,RR}\,,\nonumber\\[0.2cm]
\dot{L}_{\substack{Nu\\prwt}}^{S,RR}&=\left[-6g_s^2C_F-\frac{24}{9} e^2\right]L_{\substack{Nu\\prwt}}^{S,RR}\,,
\hspace{2.5cm} \dot{L}_{\substack{Nd\\prwt}}^{S,RR}=\left[-6g_s^2C_F-\frac{6}{9} e^2\right]L_{\substack{Nd\\prwt}}^{S,RR}\,,
\nonumber\\[0.2cm]
\dot{L}_{\substack{\nu u\\prwt}}^{S,LR}&=\left[-6g_s^2C_F-\frac{24}{9} e^2\right]L_{\substack{\nu u\\prwt}}^{S,LR}\,,
\hspace{2.5cm} \dot{L}_{\substack{\nu d\\prwt}}^{S,LR}=\left[-6g_s^2C_F-\frac{6}{9} e^2\right]L_{\substack{\nu d\\prwt}}^{S,LR}\,,\nonumber\\[0.2cm]
\dot{L}_{\substack{uN\\prwt}}^{S,LR}&=\left[-6g_s^2C_F-\frac{24}{9} e^2\right]L_{\substack{uN\\prwt}}^{S,LR}\,,
\hspace{2.5cm} \dot{L}_{\substack{dN\\prwt}}^{S,LR}=\left[-6g_s^2C_F-\frac{6}{9} e^2\right]L_{\substack{dN\\prwt}}^{S,LR}\,.
\end{align}

\section{The UV completions giving rise to tensor four-fermion operators}\label{app:UV}

The tensor four-fermion operators can stem from the UV completions of which possess either genuine and/or effective tensor interactions with both left- and right-handed neutrinos. The latter can be induced from the Fierz transformation of scalar, pseudo-scalar, and tensor operators, but not of vector and axial-vector operators. Here we list two UV complete examples that can induce the tensor four-fermion operators. 

To give rise to the neutrino-lepton tensor operators, we quote the minimal left-right symmetric model~\cite{Pati:1974yy,Mohapatra:1974gc,Senjanovic:1975rk}, which respects to $SU(2)_L\times SU(2)_R\times U(1)_{B-L}$ gauge symmetries. The Lagrangian for the Yukawa interactions in the leptonic sector reads,
\begin{align}\label{eq:MLRSM}
\mathcal{L}_\Phi\supset-\bar{L}_{Li}\left(y_{\ell ij}\Phi+\tilde{y}_{\ell ij}\tilde{\Phi}\right)L_{Rj}+\mathrm{h.c.}\,,
\end{align}
where $y_{\ell ij}$ and $\tilde{y}_{\ell ij}$ are the Yukawa couplings with generation indices $i,j=1,2,3$, $L_L\sim(\textbf{2},\textbf{1},-1)$ and $L_R\sim(\textbf{1},\textbf{2},-1)$ denote respectively for the left- and right-handed lepton doublets, and $\Phi\sim(\textbf{2},\textbf{2},0)$ is the scalar bidoublet:
\begin{align}
L_L=\begin{pmatrix}
\nu_L\\\ell_L
\end{pmatrix}\,,\quad L_R=\begin{pmatrix}
N_R\\\ell_R
\end{pmatrix}\,,\quad \Phi=\begin{pmatrix}
\phi_1^0 & \phi_1^+\\ 
\phi_2^- & \phi_2^0
\end{pmatrix}\,,\quad \tilde{\Phi}=\sigma_2\Phi^\ast\sigma_2\,.
\end{align}
Rewriting Eq.~\eqref{eq:MLRSM} in components yields
\begin{align}\label{eq:Phiterms}
\mathcal{L}_\Phi\supset-y_{\ell ij}\bar{\nu}_{Li}\ell_{Rj}\phi_1^++\tilde{y}_{\ell ij}\bar{\ell}_{Li}N_{Rj}\phi_1^--y_{\ell ij}\bar{\ell}_{Li}N_{Rj}\phi_2^-+\tilde{y}_{\ell ij}\bar{\nu}_{Li}\ell_{Rj}\phi_2^++\mathrm{h.c.}\,.
\end{align}
With Eq.~\eqref{eq:Phiterms} we can construct the tree-level leptonic process via exchange $\phi_i^\pm$ ($i=1,2$). For small momentum transfer the heavy $\phi_i^\pm$ can be integrated out, then we arrive at the effective interactions,
\begin{align}
\mathcal{L}_\mathrm{eff}=-\frac{y_{\ell pt} \tilde{y}_{\ell wr}}{M_{\phi}^2}(\bar{\nu}_{Lp}\ell_{Rt})(\bar{\ell}_{Lw}N_{Rr})+\mathrm{h.c.}\,,
\end{align}
which after the Fierz transformation becames
\begin{align}\label{eq:MLRSMeff}
\mathcal{L}_\mathrm{eff}=\frac{y_{\ell pt} \tilde{y}_{\ell wr}}{M_{\phi}^2}\left[\frac{1}{2}(\bar{\nu}_{Lp}N_{Rr})(\bar{\ell}_{Lw}\ell_{Rt})+\frac{1}{8}(\bar{\nu}_{Lp}\sigma_{\mu\nu}N_{Rr})(\bar{\ell}_{Lw}\sigma^{\mu\nu}\ell_{Rt})\right]+\mathrm{h.c.}\,.
\end{align}
Consequently, matching the neutrino-lepton tensor operator in Eq.~\eqref{eq:MLRSMeff} onto the corresponding one $\mathcal{O}_{\substack{\nu Ne\\prwt}}^{T,RR}$ in LNEFT yields
\begin{align}
L_{\substack{\nu Ne\\prwt}}^{T,RR}=\frac{y_{\ell pt} \tilde{y}_{\ell wr}}{8M_{\phi}^2}\,.
\end{align}
Note that similar discussion on this aspect can also be found in Ref.~\cite{Xu:2019dxe}.

As for neutrino-quark tensor operators, one of the feasible UV completions is to introduce a scalar leptoquark $\tilde{R}_2$~\cite{Dorsner:2016wpm}, which transforms as $(\textbf{3},\textbf{2},1/6)$ under the SM gauge group $SU(3)_C\times SU(2)_L\times U(1)_Y$. The interactions of $\tilde{R}_2$ with leptons and quarks are described by the Lagrangian
\begin{align}\label{eq:R2}
\mathcal{L}_{\tilde{R}_2}\supset-\tilde{y}_{2ij}^{RL}\bar{d}_{Ri}\tilde{R}_2i\sigma_2L_{Lj}+\tilde{y}_{2ij}^{\overline{RL}}\bar{Q}_{Li}\tilde{R}_2N_{Rj}+\mathrm{h.c.}
\end{align}
where $Q_L$ and $d_R$ stand respectively for the $SU(2)$ quark doublets and down-type quark singlets, $\tilde{y}_{2ij}^{RL}$ and $\tilde{y}_{2ij}^{\overline{RL}}$ are the Yukawa coupling matrices with $i,j=1,2,3$ stand for the generation indices. Switching to the mass basis for quarks and leptons, Eq.~\eqref{eq:R2} becomes
\begin{align}\label{eq:R2mass}
\mathcal{L}_{\tilde{R}_2}\supset&-\tilde{y}_{2ij}^{RL}\bar{d}_{Ri}\ell_{Lj}\tilde{R}_2^{2/3}+(\tilde{y}_{2}^{RL}U)_{ij}\bar{d}_{Ri}\nu_{Lj}\tilde{R}_2^{-1/3}\nonumber\\
&+(V\tilde{y}_{2}^{\overline{RL}})_{ij}\bar{u}_{Li}N_{Rj}\tilde{R}_2^{2/3}+\tilde{y}_{2ij}^{\overline{RL}}\bar{d}_{Li}N_{Rj}\tilde{R}_2^{-1/3}+\mathrm{h.c.}\,,
\end{align}
where $\tilde{R}_2^{2/3}$ and $\tilde{R}_2^{-1/3}$ are the up- and down-component of $\tilde{R}_2$ doublet, and $V$ and $U$ denote the Cabibbo-Kobayashi-Maskawa (CKM) mixing matrix and Pontecorvo-Maki-Nakagawa-Sakata (PMNS) unitary mixing matrix, respectively. Following the similar steps as done in the minimal left-right symmetric model, i.e., constructing the tree-level neutrino-quark process via exchange $\tilde{R}_2^{\pm1/3}$, integrating out the heavy degree, and making a Fierz transformation, one finally obtain the following effective interactions:
\begin{align}\label{eq:R2eff}
\mathcal{L}_\mathrm{eff}=-\frac{(\tilde{y}_{2}^{RL}U)_{wr}\tilde{y}_{2tp}^{\overline{RL}\ast}}{M_{\tilde{R}_2}^2}\left[\frac{1}{2}(\bar{\nu}_{Lp}N_{Rr})(\bar{d}_{Lw}d_{Rt})+\frac{1}{8}(\bar{\nu}_{Lp}\sigma^{\mu\nu}N_{Rr})(\bar{d}_{Lw}\sigma_{\mu\nu}d_{Rt})\right]+\mathrm{h.c.}\,.
\end{align}
Matching the neutrino-quark tensor operator in Eq.~\eqref{eq:R2eff} onto the corresponding one $\mathcal{O}_{\substack{\nu Nd\\prwt}}^{T,RR}$ in LNEFT, one obtains
\begin{align}
L_{\substack{\nu Nd\\prwt}}^{T,RR}=-\frac{(\tilde{y}_{2}^{RL}U)_{wr}\tilde{y}_{2tp}^{\overline{RL}\ast}}{8M_{\tilde{R}_2}^2}\,.
\end{align}
Note that with the similar steps, one can also obtain neutrino-quark tensor operators from another distinct scalar leptoquark $S_1\sim(\bar{\textbf{3}},\textbf{1},1/6)$, which is usually employed to simultaneously solve the anomalies in $B$ physics and muon $g-2$, see e.g., Ref.~\cite{Bauer:2015knc}.

\bibliographystyle{JHEP}
\bibliography{reference}

\end{document}